\documentclass{aastex}          
\usepackage{spr-astr-addons}    
\usepackage{url}
\usepackage{graphicx}

\begin{document}

\title{An Updated Wing TiO Sensitive Index  for Classification of M-Type Stars}

\shorttitle{<An updated Wing Index>}
\shortauthors{<F. Azizi et al.>}

\author{F. Azizi\altaffilmark{}}
\and
\author{M. T. Mirtorabi\altaffilmark{}}
\affil{f.azizi@alzahra.ac.ir}
\email{f.azizi@alzahra.ac.ir} 

\altaffiltext{}{ Department of Physics, Alzahra University, 1993891176, Tehran, Iran.}

\begin{abstract}
By careful searching of synthetic and observed spectra in a sample of
cool giant and supergiant stars, we have updated the continuum
band-passes of near-infrared Wing three filter system. This photometric
system measures the strength of Titanium Oxide (TiO) absorption in Near-Infrared (NIR) at 719 nm.
We show that new reference continuum band-passes are essentially free from molecular
absorptions and the updated TiO-index defines the temperature variation
in a sample of cool giants with less scatter. A TiO-index vs. effective
temperature calibration is derived based on new continuum band-passes.
\end{abstract}


%
\section{Introduction}
Among many absorption features in spectra of cool stars, TiO molecular absorptions are be recognized as an indicator of basic stellar parameters mainly in two type of stars, cool pulsating giants and super giants and chromospheric active stars which TiO absorption can be used as a starspot indicator \citep{one96,nef95,one98}.
The temperature sensitivity of TiO bands at near-infrared is so pronounced that these bands have always been used as a primary indicator to divide M stars into their subtypes.
Another interesting observed fact is that the TiO absorption strength is independent  of luminosity class, which is approximately the same for dwarf, giants and supergiants \citep{win78}.

\cite{ram81} defined an index by measuring the TiO$\left ( \gamma ,0,0 \right )$ absorption band strength at $\lambda$ = 886 nm in M giants, and calibrated it versus temperature by using the relation between spectral type and temperature given by Ridgway \citep{rid80}.
As shown in Figure 2 of Ramsey, the depth of the TiO band at 886 nm is a sensitive temperature indicator for giants in the range $3250\leq T_{eff}\leq 3730$ K. It is also somewhat useful in the range $3730\leq T_{eff}\leq 4000$ K. At $T_{eff}>4000$ K, the TiO bands become too weak to be useful.

\cite{win92} has invented a TiO sensitive index which was proposed to measure the absorption strength of TiO molecules at 719 nm. He showed that this index can distinguish between  M super giant sub-types. \cite{mir03} used Wing TiO-index to measure longitudinal distribution of starspots on the surface of the G8III star $\lambda$  Andromeda and have estimated a period of 10-15 years for chromospheric activity of this star.

To provide a correct measure of absorption at 719 nm Wing has estimated the continuum over  this absorption band by extrapolating the continuum in 754 and 1024 which he claimed are mostly free of absorption. This has made a three filter system which is known as the Wing filter system.

 One of the major advantages of Wing TiO-index has been its strong sensitivity  to effective temperature for atmospheres  cooler than 4000 K. The linear reduction of the index  by increasing $T_{eff}$ has made it a temperature index which has been used by several authors including himself.  The possible contamination of other molecular absorptions in continuum regions would make this linear relation inaccurate and misleading.

 Technological advancements in spectroscopic facilities which provide us with ultra high resolution spectra especially in near infrared have made it possible to better investigate the Wing continuum bands and to check if they are contaminated or if it is possible to shift them to represent a more accurate continuum, free of molecular absorbtion. The basic motivation of this paper is investigating the synthetic and also high resolution observed spectra of a sample  of cool giants to find a more precise temperature calibration for cool stars.\\

 In this paper, we present an updated TiO-index  which is proposed to measure absorption strength of TiO bands at  $\lambda$ = 719 nm, by using the high-resolution spectroscopic data from the \objectname{UVES spectrograph}. We first search in a grid of synthetic spectra to find a spectral band which closely resembles the continuum. Section 2 describes Wing system properties for measuring the TiO$\lambda$719 nm absorption strength and the definition for TiO-index. In Section 3, a modified continuum band is selected which we claim is less affected from absorption line contamination. The stellar sample is described in Section 4.
Section 5 presents the results, including  Wing TiO-index and our index values for a sample of stars along with their calibration as a function of the effective temperature.

\section{Wing System}

%
\begin{table}
\caption{Wing Three-Color Near-IR Filters} 
 \label{tbl:1}
 \begin{tabular}{@{}cccc@{}}
 \tableline  
Filter     &  Region Measured     & Central  & Band-pass  \\
              &                &Wavelength &FWHM \\
    &        &(nm) &(nm) \\
\hline A...  & TiO$\lambda$719 nm band &719 &11 \\
B...  &Continuum &754 &11 \\
C...  &Continuum &1024 &42 \\
\tableline 
 \end{tabular}
 \end{table}

\subsection{Wing Filters}
The Wing three color system \citep{win92} consists of three filters, designated A, B and C, that can be used to measure the four principal properties of cool stars: their magnitude in near-infrared region, color index in near infrared, temperature, and spectral type.

The central wavelengths and full width at half-maximum band-pass at these filters are listed in Table \ref{tbl:1}. The central wavelength of the A filter (full width at half-maximum is 11 nm) at 719 nm is one of the strongest absorption bands of TiO $\left ( \gamma ,0,0 \right )$ molecules. The B filter with central wavelength of 754 nm measures the continuum intensity in the right wing of TiO absorption band and the C filter (full width at half-maximum is 42 nm) with central wavelength of 1024 nm is located in a continuum region far from A and B.

\subsection{TiO-Index}
The strength of the TiO absorption at 719 nm is measured by dividing the stellar flux received through filter A to filter B. However, this ratio is slightly sensitive to the slope of the energy distribution of the star. \cite{win92} has removed this effect by linearly extrapolating the B and C band-passes to the center of A band and by introducing a measure of absorption at A band by a TiO index:

\begin{equation}\label{eqn:1}
TiO-Index= A - B + 0.13(B - C)
\end{equation}

where A, B and C are the apparent magnitudes in A, B and C filters. The numerical coefficient of 0.13 is measured by the spacings of the filters in wavelength.

\section{The New NIR Continua Filters}

\subsection{New Filters}
 As mentioned earlier, TiO-index is a measure of the depth  of TiO$\lambda$719 nm molecular absorption band. Precise measurement of the continuum at A band-pass will give a more accurate  absorption strength and by the way a more precise calibration. We use a set of synthetic spectra to search for more appropriate B and C bands, free from contamination of other absorption features, using Kurucz model atmosphere \citep{kur91, cast20}.

\begin{figure*}[t]
\includegraphics[width=17cm,height=15cm,natwidth=1000,natheight=600]{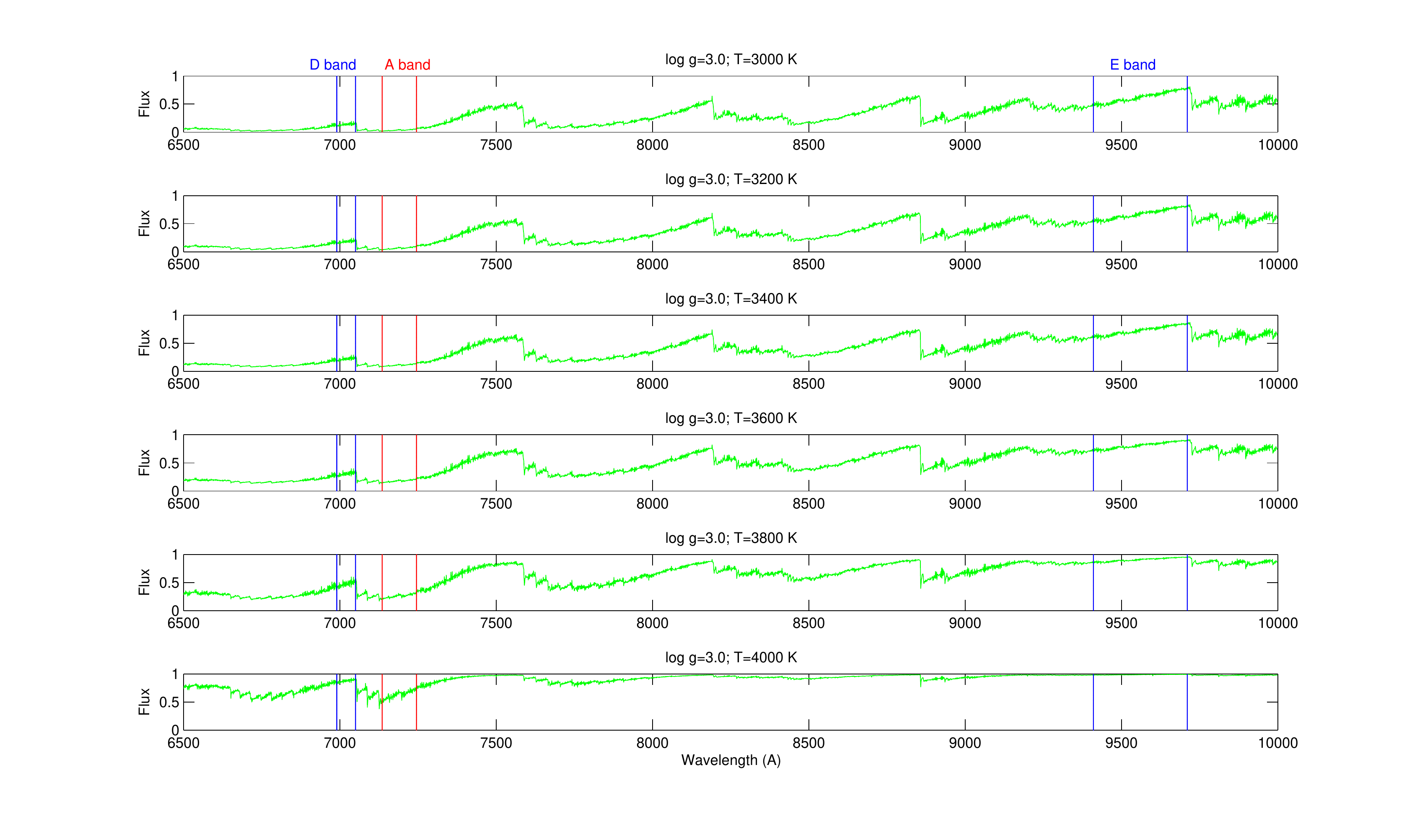}
\caption{The synthetic spectra of near-infrared absorption features of TiO$\lambda$719 nm molecule in the atmosphere of cool stars. D and E are our selected new continuum filters.}
\label{fig:1}
\end{figure*}

 With [M/H] = 0 (i.e., star abundance), $\log$ g between 3 to 5 , $\Delta \log$ g = 0.5 and effective temperature between 2900 and 4000 K, $\Delta$T = 100 K where TiO molecules have significant absorbtion at near-infrared, a number of 60 synthetic spectra were obtained. Figure \ref{fig:1} shows a few sample synthetic spectra which were produced in this exercise.

%

The results of this investigation lead to a redefinition of two new filters in the cleaner continuum areas that are designated D and E. The central wavelength and their bandwidth along with the A filter features are given in Table  \ref{tbl:2}.

The cool red stars' spectra are dominated by molecular absorption bands, especially TiO and vanadium oxide (VO) \citep{car20}.
Careful investigation of the Wing continuum shows that there are weaker molecular absorption features like TiO$\lambda$758.9 nm, TiO$\lambda$1014.7 nm, TiO$\lambda$1026.1 nm, TiO$\lambda$1003.4 nm, and VO$\lambda$753.4 nm in his selected continuum region \citep{val98, kik91}.

\begin{table*}
\caption{The New Near-IR Filters}\label{tbl:2}
\centering\begin{tabular}{ @{}cccc@{} }
\tableline
   Filter     &     Region Measured       & Central  & Band-pass  \\
                &                &Wavelength &FWHM \\
 & &(nm) &(nm) \\
\tableline A... & TiO$\lambda$719 nm band &719 &11 \\
D... &Continuum &702 &6 \\
E... &Continuum &956 &30 \\
\tableline
\end{tabular}
\end{table*}

 \citet{kik91} carried out a survey of a sequence of M and K stars' spectra in the red and infrared ranges, have provided a reference list for the absorption features in these stars of all luminosity classes. In Table \ref{tbl:3}, VO absorption wavelength regions are merely given  for a quick review.
Also, it is remarkable that the dependence of VO absorption with spectral type in the dwarfs and giants is similar \citep{one04}.\\
Table \ref{tbl:3} shows that the new continuum band-passes D and E are essentially free of molecular absorption.

%
\begin{table*}
\caption{VO absorption features found in late-K to late-M stars spectra from 6300 to 9000 $\AA$; from  \citet{kik91}}\label{tbl:3}
\centering\begin{tabular}{@{}l l l@{}}
\tableline
 Wavelength   &  Absorption   &  Notes \\
(nm)  & Feature&         \\
\tableline 733.4, 737.2, 739.3, 740.5, 741.8, &VO &Seen only in late-M stars  \\
 747.2, 753.4 &        &            \\
\\
785.1, 786.5, 789.7, 790.0, 791.9,  &VO &Obvious in late-M dwarfs  \\
792.9, 793.9, 796.1, 796.7, 797.3 &              \\
\\
852.1, 853.8, 857.4, 859.7, 860.5,  &VO &Found only in late-M stars  \\
862.4, 864.9, 866.8 &                       \\
\\
\tableline
\end{tabular}
\end{table*}

\subsection{The Updated TiO-Index}
Using the new filters, the TiO-index is redefined as

\begin{equation}
TiO-Index= A - D+ 0.016 (D-E)
\end{equation}

where A, D and E are the apparent magnitudes in A, D and E filters. According to the central wavelength of the new filters, the numerical coefficient of 0.016 is measured by the spacings of the filters in wavelength.

\section{The Data}
To check the feasibility of the new TiO-index, we select a set of 23 giant K5-M8 observed spectra from UVES Paranal Observatory Project \citep{jeh05,bag03}. All target stars have been observed by the VLT UT2 (Very Large Telescope) at European Southern Observatory using the Ultraviolet-Visual Echelle Spectrograph (UVES). UVES is a very efficient high resolution spectrograph designed to operate from about 300 to 1100 nm.\footnote{The UVES webpages: http://www.eso.org/instruments/uves/} 
The spectra are accessible from the ESO archives.\footnote{The ESO UVES archive: http://archive.eso.org/wdb/wdb/eso/\\uves/form/}
These sample stars are listed in the first column of Table \ref{tbl:4}.

%
\begin{table*}[t]
\caption{TiO-Index and $T_{eff}$ of sample stars;Wing and updated indices }\label{tbl:4}
\centering\begin{tabular}{@{}l l l l l l l@{}}
\tableline
 Star        & Spectral  & Obs. & Wing  & Updated &$T_{eff}$   &   Ref., $T_{eff}$  \\
                & Type        &  Date & TiO-Index & TiO-Index &   &                      \\
\tableline HD11695 &M4 III &2003-02-08 &0.728 &0.749 &3550 & 16 \\
HD18191 &M6 III &2001-12-05 &1.833&1.092 &3266 & 4,12,13 \\
HD18242 &M7 III &2001-07-24 &1.886&1.397 &3102 & 3,9  \\
HD18884 &M1.5 III &2003-01-02 &0.174&0.362 &3788 & 6,8,14,17 \\
HD29139 &K5 III &2002-03-04 &0.024 &0.279&3943 &  2 \\
HD29712 & M8 III & 2002-12-28 & 1.535&1.967 & 2900 & 1,5 \\
HD39801 &M2 Iab &2002-09-13 &0.307 & 0.403 &3659& 12 \\
HD102212 &M1 III &2004-03-10 &0.050&0.311 &3824 & 11,14,15 \\
HD118767 &M5 III &2001-07-09 &1.382&1.036 &3339 & 3,9 \\
HD119149 &M2 III &2001-07-12 &0.243&0.406 &3782 & 12 \\
HD120052 &M2 III &2001-07-13 &0.207&0.397 &3715 & 12,14 \\
HD123934 &M1 III &2002-01-24 &0.090&0.320 &3850 & 12 \\
HD126327 &M7.5III &2006-03-20 &1.725&1.578 &3050 & 11,13 \\
HD146051 &M0.5 III &2006-03-27 &0.428 &0.282&3879 & 6,18 \\
HD149447 &K5 III &2003-03-13 &0.001&0.242 &4000& 12 \\
HD167006 &M3 III &2003-09-07 &0.403&0.537 &3640 & 12 \\
HD167818 &K5 III&2001-06-23 &0.092&0.080 &4450 & 10 \\
HD175588 &M4 II &2003-05-01 &0.815&0.746 &3550& 4 \\
HD189124 &M6 III &2001-06-20 &1.396&0.992 &3334 & 3,9 \\
HD189763 &M4 III &2001-07-28 &0.676&0.668 &3568 & 3,9 \\
HD213080 &M4.5 IIIa &2001-07-12 &0.708&0.696 &3514 & 3,9 \\
HD214952 &M5 III &2002-05-24 &0.810&0.742&3480 & 7 \\
HD224935 &M3 III &2001-11-29 &0.536&0.581&3634 & 6,12 \\
\tableline

\tablerefs{(1) \citealt{bed97}; (2) \citealt{bel89}; (3) \citealt{bel97}; (4) \citealt{cen07}; (5) \citealt{dum98}; (6) \citealt{dyc98}; (7) \citealt{eng06}; (8) \citealt{flow96}; (9) \citealt{hou00}; (10) \citealt{jon92}; (11) \citealt{kol12}; (12) \citealt{mc12}; (13) \citealt{pru11}; (14) \citealt{str00}; (15) \citealt{sun13}; (16) \citealt{wit04}; (17) \citealt{wit06} (18) \citealt{wu20}}

\end{tabular}
\end{table*}

%
\begin{figure}[t]
\includegraphics[width=1\columnwidth,natwidth=600,natheight=600]{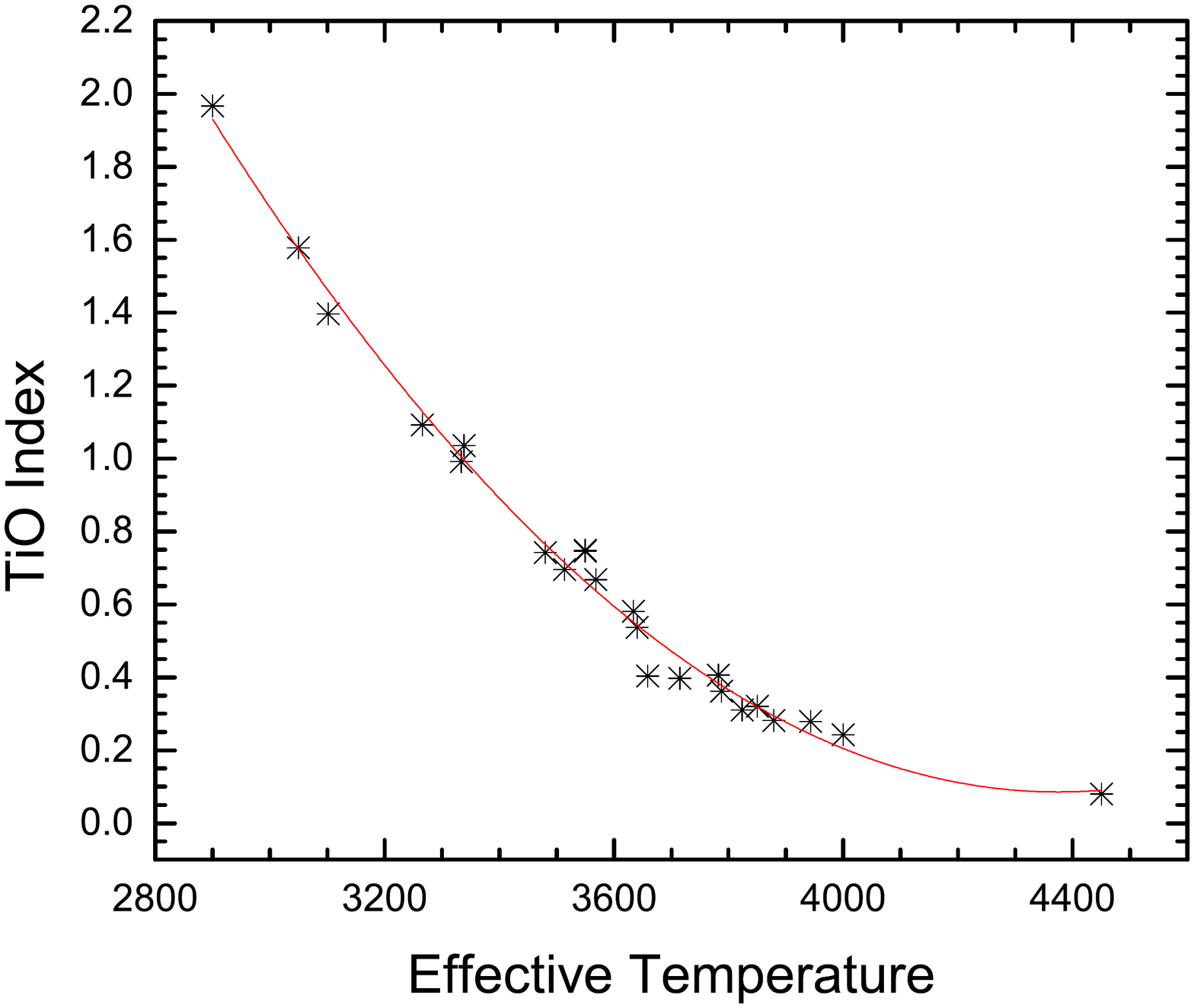}
\caption{ Calibration of TiO-index relative to $T_{eff}$  for K5-M8 stars; new filters}
\label{fig:2}
\end{figure}

%
\begin{figure}[t]
\includegraphics[width=1\columnwidth,natwidth=600,natheight=600]{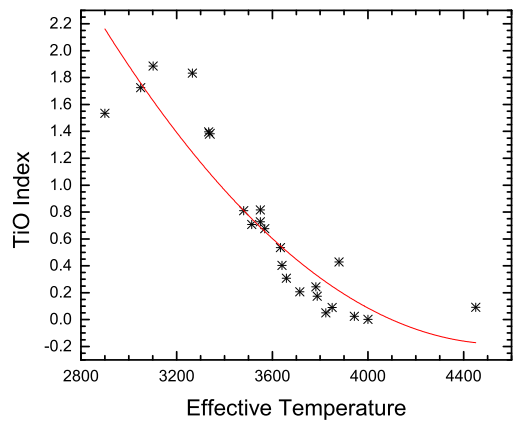}
\caption{ Calibration of TiO-index relative to  $T_{eff}$ for K5-M8 stars; Wing filters}
\label{fig:3}
\end{figure}

\section{Results}
To determine spectral types, it is necessary to calibrate the TiO-index as a function of effective temperature by our sample stars of known spectral types.
First, the TiO-index of the sample stars is calculated by using our proposed index. The results are given in the fifth column of Table \ref{tbl:4}. The effective temperatures of most of the sample stars were determined by averaging over the temperatures found from the existing literature. For stars where these were not found in the literature, we used the calibration of  Houdashelt and Bell \citep{hou00,bel97} from their published V-K indices (results are given in the sixth column of Table \ref{tbl:4}).
Figure \ref{fig:2} represents the variation of updated TiO-index relative to effective temperature for our sample stars. A quadratic least-squares fit was performed to the points as

\begin{multline}
TiO'-Index= 8.484\times 10^{-7}  T_{eff}^{2}-7.424  \times 10^{-3} T_{eff}\\ + 16.32
\end{multline}

 The dispersion of the residuals around the fit is 0.04532. As it can be seen from the curve, the value of TiO$'$-index is declining with increasing temperature and for temperatures above 4500 K, asymptotically approaching zero where the TiO absorption feature is not present or is very weak. By decreasing temperature, the TiO$'$-index increases up to a value about 2.00  which corresponds to spectral type M8.

To compare our results with original Wing calibration, we recalculate the Wing TiO-index for these stars. The results of calculation are given in the fourth column of Table \ref{tbl:4}.

Figure \ref{fig:3} represents the variation of  the original TiO-index relative to the effective temperature for our sample stars. A quadratic least-squares fit was performed to the points as

\begin{multline}
TiO-Index= 8.489\times 10^{-7}  T_{eff}^{2}- 7.745 \times 10^{-3} T_{eff}\\ + 17.48
\end{multline}

 The dispersion of the residuals around the fit is 0.2431. This curve displays the same trend in Fig. 3 but with more scatter, especially at later spectral-types.

\section{Conclusions}
In this paper, we have presented a new continuum filters set for measuring  the TiO-index and an updated calibration of the TiO-index with effective temperature. By comparing the dispersion of the residuals around the fit for the two curves, it is clear that the quality of the fit for the updated TiO-index is better than for the original Wing index. Moreover, we used the UVES high resolution spectrum. The high scatter in Fig. 3 might be due to the presence of some absorption lines in the original Wing B and C continuum band-passes. The new TiO-index proposed here can provide more precise information about starspot activity in solar like stars. To test the proposed revised Wing TiO photometric  system, actual photometry should be carried out. Another advantage of the new system is that the E filter at 956 nm is better placed at shorter wavelength than the original Wing C filter at 1024 nm. Most CCD detectors rapidly decrease in sensitivity at wavelengths $>$ 1000 nm.

%
\acknowledgment
This project is based on observations collected with the UVES spectrograph at the VLT/UT2 telescope(Paranal Observatory, European Southern Observatory, Chile, Science Archive Facility).

%
\nocite{*}
\bibliographystyle{spr-mp-nameyear-cnd}

\end{document}